\documentclass[letterpaper
,aps,pra
,twocolumn
,showkeys
,nofootinbib
]{revtex4-2}

\usepackage{graphicx}
\usepackage{dcolumn}
\usepackage{bm}
\usepackage{latexsym}
\usepackage{ifsym}
\usepackage{stix}
\usepackage[utf8]{inputenc}
\usepackage{amsmath}
\usepackage{setspace}
\usepackage{booktabs}
\usepackage{tabularx}

\begin{document}

\bibliographystyle{apsrev}


\title{Calculation of the near threshold photoabsorption for the ground and excited sodium states}

\author{Constantine E. Theodosiou}
\affiliation{Department of Physics and Astronomy, Manhattan College, Riverdale, New York 10471}
\email{E-mail: constant.theodosiou@manhattan.edu}

\date{\today}
\begin{abstract}
We revisit the photoabsorption from the ground state of Na, ending below and above the ionization threshold, with special emphasis on the shape of the photoionization cross section around the Cooper minimum.  The present treatment, using a self-consistent central field and including core polarization effects, matches and improves over the results of other recent calculations.  
The  photoabsorption from the excited $4s,$ $3p$, and $3d$ states is also calculated, along with the corresponding optical oscillator strengths.  The results are compared with all published experimental and with the most representative of the theoretical data.  The comparisons confirm the results of some of the measurements, but raise serious concerns about the validity of others. The continuing disagreement in the case of the ground state with the only one available experimental measurement, that is over 55 years old, points to the needs for new, refined, measurements.  
\end{abstract}

\pacs{...}

\maketitle

\section{Introduction}\label{} 
The ground-state photoabsorption cross sections of alkali metal atoms are known to have distinct minima called Cooper minima \cite{Cooper1962, Fano1968}, just above the ionization threshold for Na, K, Rb, and Cs, and below threshold in Li. Although the active electron in these atoms is ``hydrogenic'' in nature, i.e. a single electron outside a spherically symmetric core charge distribution, the alkali metal atoms have a finite core that perturbs the otherwise hydrogenic wavefunctions of the valence electrons  to cause enough ``out-of-phase" shift and change in the initial-final state wavefunction overlap that at the appropriate energy the photoabsorption transition matrix element goes through a sign change.  The resulting transition matrix element zeroes result in minima in the observed cross sections.
The width and depth of these minima change with the atomic number as the spin-orbit interaction also changes correspondingly \cite{Seaton1951}. Sodium has the additional interesting feature that it is the first in the periodic table to have a filled $p$-shell but does not have a strong enough core to support a $d$ orbital in its ground state which creates a type of instability in the formation of excited states.  An effect most noticeable in alkalis is the polarization of the core by the valence electron, which yields to additional correction to the theoretical treatment and needs to be treated explicitly to reproduce experimental results on energy levels and photoabsorption.  For the ground state, 3$s$, there are only two actual measurements of the photoionization cross section, by Ditchburn et al.\ \cite{Ditchburn1953} and by Hudson and Carter \cite{Hudson1965}.  Those measurements are over 55 years old and were plagued with effects by molecular photoabsorpion by other species present during the experiments.  Yet, they display the existence of the Cooper minima near the ionization threshold.  A precise measurement in the case of potassium by Sandner et al. \cite{Sandner1981}, who overcame such complications by using a time-of-flight technique to separate the atomic K signal, has helped distinguish the successful theoretical treatments. It is disappointing that no recent measurement has been attempted for sodium to date. 

Being relatively easy to calculate at a first approximation, sometimes even using modified analytic atomic potentials and wavefunctions, the alkalis have been a testing ground for many theoretical calculations. However, when it comes to accurately reproducing some experimental data, some of them become clearly more successful. In the case of sodium the treatments of the photoionization close to threshold include Hartree-Slater \cite{Fano1968}, Hartree-Fock \cite{Chang1975}, many-body perturbation \cite{Chang1975}, multi-configuration Hartree-Fock \cite{Saha1989}, relativistic random-phase approximation \cite{Fink1986}, several semi-empirical \cite{Weisheit1972, Marinescu1994}, and, more recently, configuration interaction Pauli-Fock approach with core polarization potentials \cite{Petrov1999, Petrov2000}, Dirac-based B-spline R-matrix \cite{Zatsarinny2010}, and ``fully-relativistic''\cite{Singor2021} treatments. These treatments are usually satisfactory just above threshold or away from threshold, predicting the ``general" behavior of the observed data.  The theoretical works of \cite{Petrov1999, Singor2021} reproduce the general Na($3s$) position of the minimum fairly accurately, but there are disagreements on the width of the minimum and its exact position.  The (old) experiments \cite{Ditchburn1953,Hudson1965} are not of significant guidance, as they are hampered by systematic errors attributable to other present species in their targets and thus have larger errors near the Copper minimum that mask the true magnitude of its depth.

Several recent calculations \cite{Petrov1999, Zatsarinny2010, Singor2021} have also addressed the photoabsorption from \textit{excited}  alkali metal states.  Several experimental papers \cite{Amin2008, Yar2013, Kalyar2016, Rafiq2008, Baig2007} have also studied the photoabsorption by selected (low) excited states and yielded new information on both the absorption oscillator strength distributions below the ionization threshold, as well as the photoionization cross sections near the threshold.  The comparison between experiment and theory for the excited states is not uniformly good.

In view of these comparisons, we have revisited our earlier work on alkali lifetime calculations and photoionization, and present here some refined results. Our results are also tested for accuracy via comparison with the available accurate reduced matrix elements calculated by Safronova et al.\ \cite{Safronova1998} for the transitions for which numbers have been published.  We have performed calculations for all single-excitation states of sodium for which there exist experimental measurements and we present the comparison for all of them.  From the plethora of theoretical treatments we selected representative ones that are at least as sophisticated as ours and those that yield good comparison with measurements.

\section{Method of Calculation}\label{}

A method was developed by this author \cite{CET1984a} to calculate accurate transition matrix elements for alkali-like and helium-like systems for which no substantial core excitation or channel mixing is present. The method employs a self-consistent Hartree-Kohn-Sham potential to describe the atomic field supplemented by {\it ab initio} core-polarizabilities and has been successful in predicting accurate oscillator strengths and excited state lifetimes for a variety of atomic systems 
(e.g., see  \cite{CET1984b, Curtis1993,  CET1996,  CET2022a, CET2022b} and references therein). The general approach is refined here to account for more salient features of the transitions at hand, especially the atomic core radius parameter used.

The wave functions are obtained by solving the Schr\"odinger equation
\begin{equation}
\left[ \frac{d^2}{dr^2}-V(r)-\frac{l(l+1)}{r^2}+E_{nl}\right]P_{nl}(r)=0.
\end{equation}
The central potential
\begin{equation}
V(r)=V_{{\rm HKS}}(r)+V_{{\rm pol}}(r)+V_{{\rm so}}(r)=V_{{\rm m}}(r)+V_{{\rm so}}(r)
\end{equation}
consists of three terms: $V_{\rm HKS}(r)$, a Hartree-Kohn-Sham-type \cite{Desclaux1969} self-consistent field term,
\begin{equation}
V_{\rm pol}(r)=-\frac 12\frac{\alpha _d}{r^4}\left\{ 1-{\rm exp}\left[-(r/r_c)^6\right] \right\}                             
\end{equation}
a core-polarization term, and
\begin{equation}
V_{{\rm so}}(r)=-\frac 12\alpha ^2\left\{ 1+\frac{\alpha ^2}4[E-V(r)]\right\} ^{-2}\frac 1r\frac{dV_{\rm m}(r)}{dr}\vec L\cdot \vec S
\end{equation}
a spin-orbit interaction, Pauli approximation term. Here $\alpha_d$ and $ \alpha_q$ are the dipole and quadrupole polarizabilities of the core \cite{Johnson1981}, $r_c$ is a cut-off distance and $\alpha$ is the fine-structure constant.

The necessary radial matrix elements were calculated using the modified dipole operator expression similar to the one used by Norcross \cite{Norcross1973},
\begin{widetext}
\begin{equation}
\begin{aligned}
R(nl,n'l'j')=\left< n'l'j' \left| r  \left[ 1-\frac{\alpha _d}{r^3} \left( 1-\frac{1}{2} \left[ \text{exp} \left\{ -(r/r_{cl})^3 \right\} 
+\text{exp}\left\{-(r/r_{cl^\prime})^3\right\}  \right] \right) \right] \right| nlj \right>.
\end{aligned}
\end{equation}
\end{widetext}

The cutoff distances $r_{cl}$ are taken to be equal to the values used in the polarization potential $V_{pol}$ needed to reproduce the lowest experimental energy for each symmetry; they are different for each value of $l$.  They are the only adjustable parameters in this approach.

The photoabsorption cross section is obtained using\cite{Sobelman1992}
\begin{equation}
\sigma(nlj\rightarrow n'l'j')=\frac43 \pi^2a_0^2\alpha (\epsilon_{n'l'j'}-\epsilon_{nlj})\frac{1}{2j+1} S(nlj,n'l'j')
\end{equation}
where energy is given in Rydbergs, and the absorption oscillator strength $f$ is 
\begin{equation}
f(nlj\rightarrow n'l'j')=\frac23(\epsilon_{n'l'j'}-\epsilon_{nlj}) \frac{1}{2j+1}S(nlj,n'l'j')
\end{equation}
The line strength $S(nlj,n'l'j')$ is also known as the square of the reduced dipole matrix element  and in the case of alkali metal atoms takes the form
\begin{equation}
\begin{aligned}
S(nlj,n'l'j')=&\text{max}(l,l')(2j+1)(2j'+1)\\
&\times
\begin{Bmatrix}  
l & l' & 1 \\
j' & j & \frac{1}{2} \\
\end{Bmatrix}
^2
R(nlj,n'l'j')^2.
\end{aligned}
\end{equation}

When the final state is in the continuum, $\left|n'l'j'\right>$ is replaced by $\left|\epsilon l'j'\right>$ and the total photoionization cross section of an initial state $\left|nlj\right>$ to the continuum comprises from the sum of the partial cross sections for a photon energy $E=\epsilon-\epsilon_{nlj}$:
\begin{equation}
\begin{aligned}
\sigma_{nlj}(E)=&\frac43 \pi^2a_0^2\alpha (\epsilon-\epsilon_{nlj}) \sum_{l'j'}\text{max}(l,l')(2j'+1)\\
&\times
\begin{Bmatrix}  
l & l' & 1 \\
j' & j & \frac{1}{2} \\
\end{Bmatrix}
^2
R(nlj,\epsilon l'j')^2.
\end{aligned}
\end{equation}

The continuum photoionization cross section 
joins smo-othly at the energy threshold with the oscillator strength distribution for discrete states using the formula
\begin{equation}
\sigma(nlj\rightarrow{}n'l'j')=\frac{2\pi^2\alpha\hbar^2}{m} \frac{df}{dE}\\
\end{equation}
where
\begin{equation}
\frac{df}{dE}\equiv \frac{df_{n'l'j'}}{d\epsilon_{n'l'j'}} =\frac{1}{2}(n_{j'}^*)^3 f(nlj\rightarrow n'l'j').
\end{equation}

Our approach uses relaxed, $l$-dependent central potentials within the Hartree-Kohn-Sham (HKS) approximation.  The HKS versus Hartree-Slater (HS) approach was chosen because first, it satisfies the virial theorem, and second, produces a more compact Na$^+$ core that makes the calculation of core cutoff radii more feasible for all involved states.  Thus the use of different ionic core potentials for the initial and final states introduces a ``relaxation'' multiplicative factor $N=\left<1s|1s'\right>^2\left<2s|2s'\right>^2\left<2p|2p'\right>^6$ in the matrix elements, which for Na was found to be negligibly different than unity (0.99940 for the case $4f$).   

\section{Results and Discussion}

\subsection{Na $3s$ state}

\begin{figure}[h]\centering\small\label{Na_cs(3s)}
\includegraphics[width=8.59cm]{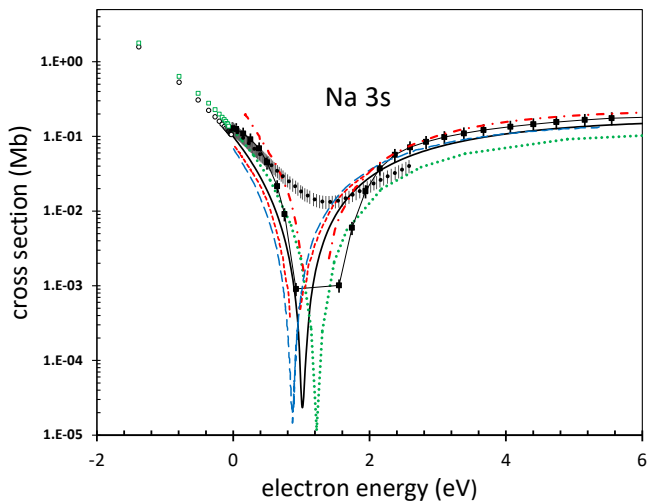}
\caption{\label{fig: Na_cs(3s)} (Color online) Na ($3s$) photoabsorption cross section around the first threshold.
\textit{Experimental:} $\mdwhtsquare$, Hudson and Carter\cite{Hudson1965}, $\mdblkcircle$, Ditchburn \cite{Ditchburn1953}.
\textit{Theoretical:} dotted line ( green), Weisheit\cite{Weisheit1972};  short dashed line (red), Petrov et al.\ \cite{Petrov1999}; long dashed line (blue), Singor et al.\ \cite{Singor2021}; dash-dot line (red), Konovalov and Ipatov \cite{Konovalov2016}; Black line, present results.  The respective data points below the threshold represent the optical oscillator strengths.}
\end{figure}

Fig.\ \ref{fig: Na_cs(3s)} displays the results of our calculations for the Na $3s$ photoabsorption.  They are compared with the theoretical results of Weisheit \cite{Weisheit1972} using a semiempirical potential approach, the work of Petrov et al.\ \cite{Petrov1999} using the variational principle to compute the core polarization potential, and the the work of Singor et al.\ \cite{Singor2021} using a fully relativistic approach. All the theoretical works need to use a cutoff radius for the core polarization potential. They all show a deep (Cooper) minimum but do not fully agree on its location in energy.  It is important to note that all these four treatments utilize as input atomic polarizabilities and atomic energy levels and adjust the dipole core-polarization potential cutoff radius to reproduce the experimental energy levels.  The semiempirical calculations of Weisheit, have also been adjusted to reproduce the location of the experimentally observed minimum. 
Table I gives a comparison of the various calculated values for the location and depth of the minimum, along with the value of the photoionization cross section at threshold, and compares them to the available experimental values.   

\begin{table}[ht]\footnotesize\centering\label{Table0}
\caption{Comparison of the measured and calculated values for threshold value $\sigma_\text{thr}$, Cooper minimum location $\epsilon_\text{min}$, and Cooper minimum value $\sigma_\text{min}$ for the Na ($3s$) state.  $\alpha_d(0)$ and $\alpha_d(\omega)$ indicate the present results using static and dynamic polarizabilities, respectively. }
\begin{tabular}{cccccccc}
\toprule\toprule
&Exp. \cite{Hudson1965, Ditchburn1953} &${\ \ }\alpha_d(0)$&{\ \ }$\alpha_d(\omega)$ &\cite{Weisheit1972} & \cite{Aymar1978}&\cite{Petrov1999} & \cite{Singor2021} \\ 
\midrule
$\sigma_\text{thr}$(Mb)   &{\ \ }0.129  &{\ \ }0.107  &{\ \ }0.105  &{\ \  }0.1198  &{\ \  }0.143   &  {\ \  }0.075   &{\ \  }0.08\\  
\midrule
$\epsilon_\text{min}$(eV)&{\ \ }1.0-1.4&{\ \ }1.041  &{\ \  }1.068  &{\ \  }1.252   &{\ \  }1.224   &{\ \  }0.92  &{\ \  }0.87\\ 
\midrule
$\sigma_\text{min}$(Mb)&{\ \ }<0.001&{\ \ }1.8E-5&{\ \  }2.3E-5&{\ \  }9.0E-6 &{\ \  }           &{\ \  }8.5E-4&{\ \  }2.5E-5\\
\bottomrule
\end{tabular}
\end{table}
The overall good agreement between the theories on the location of the minimum is remarkable. The old experimental data  of Hudson and Carter \cite{Hudson1965} are also shown for guidance; they reproduce the ``wings'' well, but are not accurate around the bottom of the minimum.  However, as we pointed out earlier, these measurements are not dependable around the minimum due to to the presence of other atomic and molecular species in the experimental setup of that work.   Nevertheless, it is helpful to examine the plausibility of the measurements above the minimum where they show an uptake as energy increases.  

It has been pointed in the literature, e.g. see \cite{CET1984a,Norcross1973}, that for the transition matrix element a more physical treatment would be to use the dynamic core polarizability $\alpha _d (\omega )$ where $\hbar \omega$ is the photon energy, instead of the static value $\alpha _d (0) \equiv \alpha _d $.  To a first approximation, the two quantities are related as
\begin{equation}
\alpha_d (\omega)\approx \alpha_d (0)/[1-(\hbar\omega/\Delta E_\text{rc})^2],
\end{equation}
where $\Delta E_\text{rc}$ is the resonance energy of the core.  As such, the approximation is valid for frequencies less than this energy.
\begin{figure}[b]\centering\small\label{Fig3a}
\includegraphics[width=8.6cm]{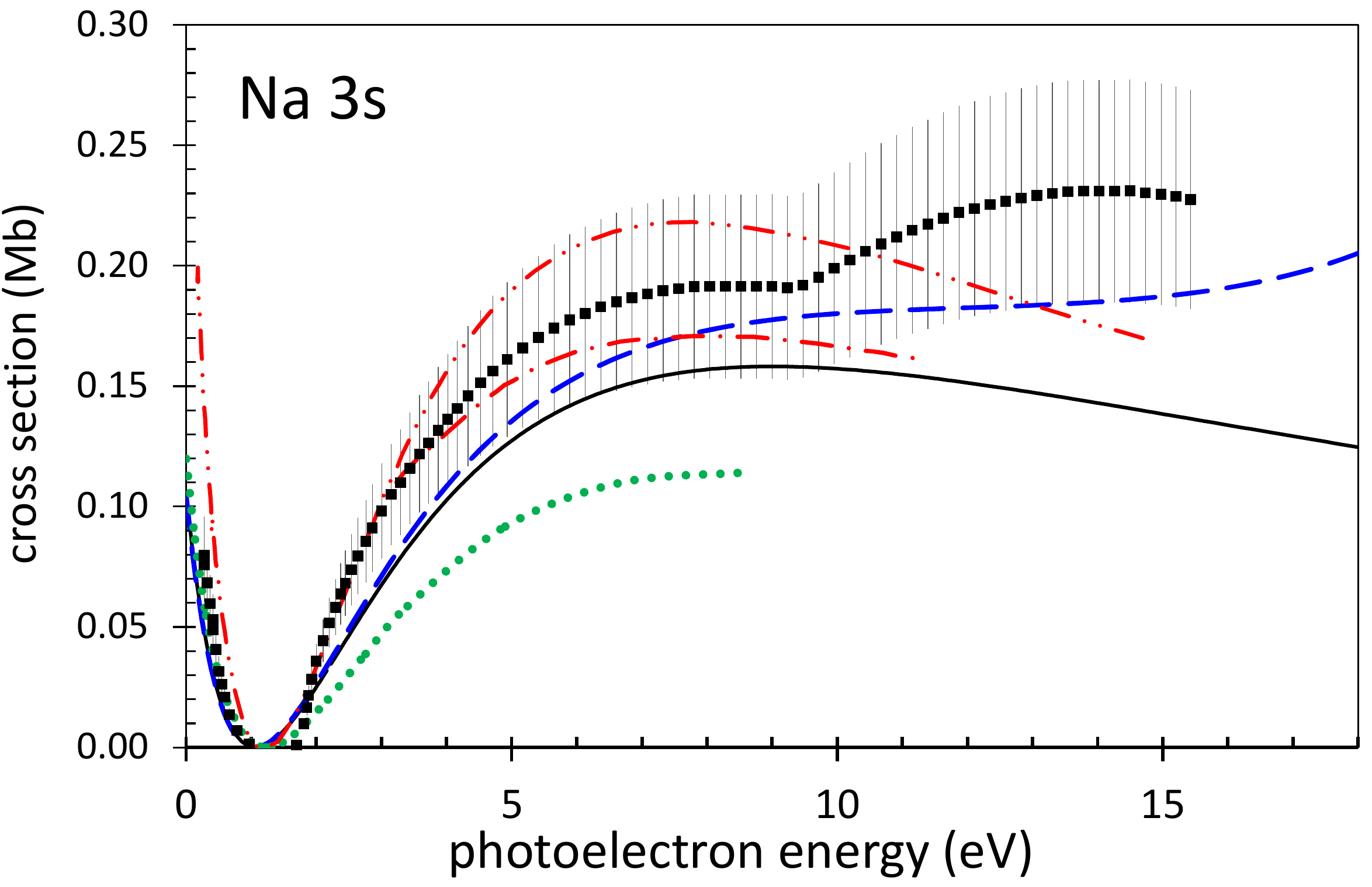}
\caption{\label{fig: Na_3s(cs)a(dyn)} (Color online) Na ($3s$) photoionization cross section below the core resonance threshold.
Experimental:  $\mdblkcircle$, Hudson and Carter\cite{Hudson1965}.
Theoretical: dash-dot, Weisheit\cite{Weisheit1972}; dash-dot line (red), Chang \cite{Chang1975}; dash-double dot line (red), Konovalov and Ipatov \cite{Konovalov2016}; solid line (black), present results with $a_d(0)$; dashed line (blue), present results with $a_d(\omega)$. }
\end{figure}

Fig.\ \ref{fig: Na_3s(cs)a(dyn)} shows that the use of the dynamic polarizability generally reproduces the upward trend of the experimental data of Hudson and Carter \cite{Hudson1965} at higher energies as one approaches the core excitation threshold of about 30.8 eV, albeit the detailed shape of the two curves is different.  Chang has given \cite{Chang1975} a first hint that many-body effects could explain the upward shift of the experimental data for increasing energies, as seen from this Figure.  A recent work by Konovalov and Ipatov\cite{Konovalov2016} includes the effects of \textit{dynamic} core polarization and as shown in Fig.\ \ref{fig: Na_3s(cs)a(dyn)}, also indicates an increase of the cross section for higher energies.  This work yields a narrower minimum; however, its threshold value of over 0.2 Mb is twice as large as  the experimental value and the projected by our accurate oscillator strengths (see Fig. \ref{fig: Na_f(3s-np)} below) value. 

It is clear from the above discussion that new experimental data are much needed to replace the currently available single set, i.e. the ones of Hudson and Carter that are over 55 years old. Accurate experimental data will help us further refine the correct theoretical approach to calculating these processes.  

Our present calculations extend below the first ionization threshold, to photoabsorption reaching the high Rydberg states.  The relevant absorption oscillator strengths are presented in Fig.\ \ref{fig: Na_f(3s-np)} and compared with available experimental and theoretical results.  The display is in the form of ${df/dE}=fn^{*3}/2$ vs.\ $n^*$ to show the asymptotic behavior as one approaches the ionization threshold.  The Weisheit \cite{Weisheit1972} results agree with ours at lower $n$ but are larger at higher $n$.  The available data from the old NBS compilation \cite{Wiese1969} are also shown; they deviate from our smooth curve for $n>11$. The latter data are over 50 years old and carry a 25-50\% estimated error. The more recent data of Nawaz et al.\ \cite{Nawaz1992} are more in line with our calculations; their values seem to scatter for $n>20$.
For absorption to the lower five n$p$ states, we are also presenting the theoretical data from McEachran et al.\ \cite{McEachran1969} and the values deduced from the matrix elements published by Safronova et al.\ \cite{Safronova1998}.  The agreement with both the latter two works is excellent and their data points are indistinguishable from ours on the graph of Fig.\ \ref{fig: Na_f(3s-np)}.  
\begin{figure}[h]\centering\small\label{Na_f(3s-np)}
\includegraphics[width=8.00cm]{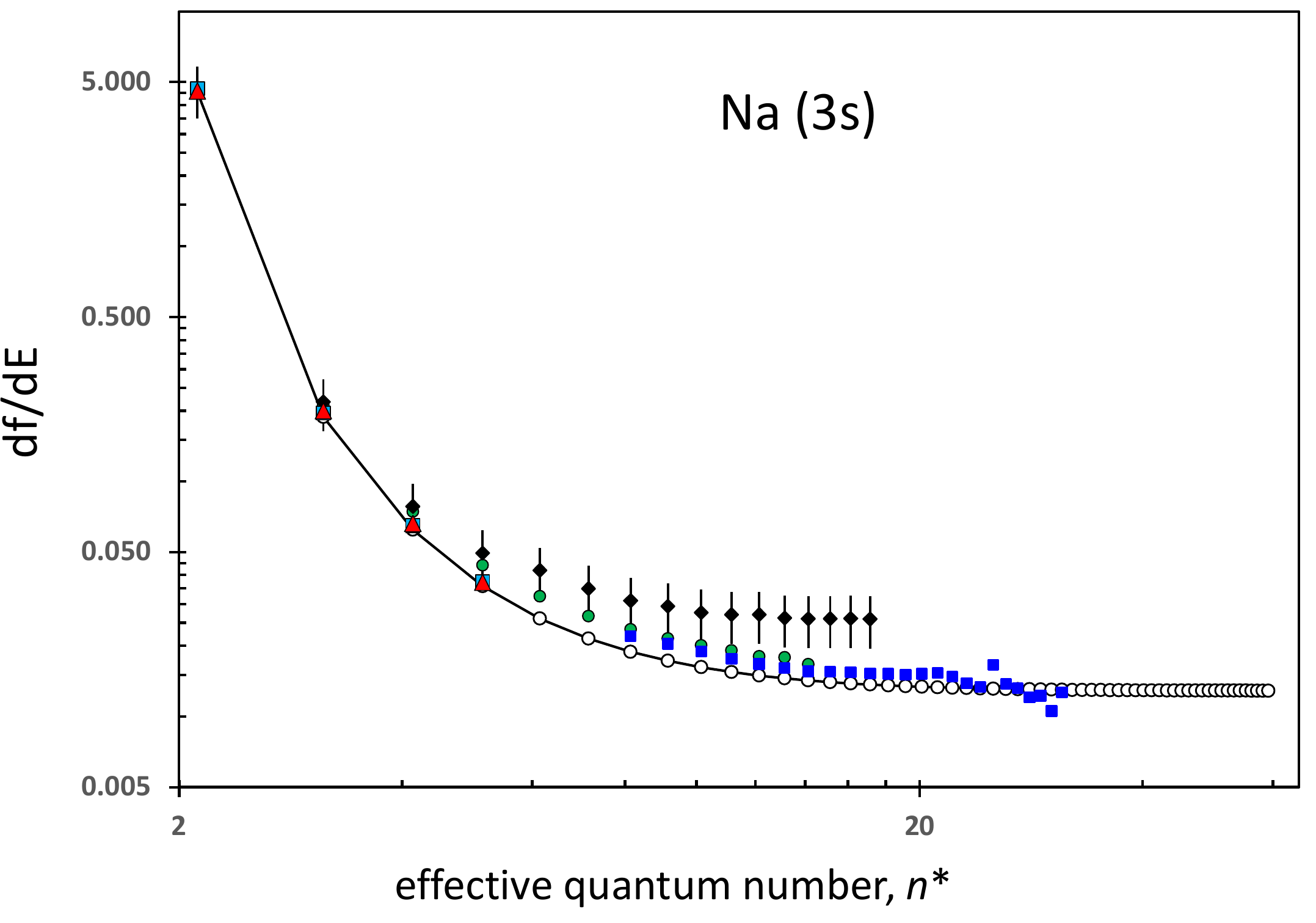}
\caption{\label{fig: Na_f(3s-np)} (Color online) $df/dE$ for  Na $3s$$\rightarrow$$np$ transitions. Experimental:
$\mdblkdiamond$, data taken from Wiese et al.\ \cite{Wiese1969}; $\mdblksquare$ (blue), Nawaz et al.\ \cite{Nawaz1992}. 
Theoretical: $\mdwhtdiamond$, McEachran et al.\ \cite{McEachran1969};  $\triangle$ (red), using the reduced matrix elements of Safronova et al.\ \cite{Safronova1998}; $\mdblkcircle$ (green), Weisheit\cite{Weisheit1972}; $\mdwhtcircle$, present. }
\end{figure}

To gauge the accuracy of our calculations, at least for lower $n$, we compare in Table II our $3p_j$$\rightarrow$$nd_{j'}$ and $3p_j$$\rightarrow$$ns$ reduced electric-dipole matrix elements (i.e. the $\sqrt{S(nlj,n'l'j')}$ in Eq.(8)) to those  published by Safronova et al.\ \cite{Safronova1998} and considered to be of high accuracy. The agreement between the two sets is within 1\% for all quoted transitions.  This comparison gives us added confidence in the overall accuracy of our calculations and the validity of the presented arguments. 
\begin{table}[h]\small
\centering\label{Table2}
\caption{Absolute values of some reduced electric-dipole matrix elements of sodium. CET, this work; SDJ, Ref.\ \cite{Safronova1998}}
\vspace{0.5mm}
\begin{tabular}{ccc}
\toprule\toprule
{Transition}&\hspace{5mm}
{CET}     &\hspace{5mm}{ SDJ}\\
\midrule
$3p_{1/2}$-$3s_{1/2}$ & \hspace{5mm}3.513 &\hspace{5mm}3.531	\\
$3p_{3/2}$-$3s_{1/2}$ &\hspace{5mm} 4.968 &\hspace{5mm}4.994	\\
$4s_{1/2}$-$3p_{1/2}$ & \hspace{5mm}3.585 &\hspace{5mm}3.576	\\
$4s_{1/2}$-$3p_{3/2}$ & \hspace{5mm}5.082 &\hspace{5mm}5.068	\\
$3d_{3/2}$-$3p_{1/2}$ &\hspace{5mm} 6.783 &\hspace{5mm}6.802  \\
$3d_{3/2}$-$3p_{3/2}$ & \hspace{5mm}3.038 &\hspace{5mm}3.046  \\
$3d_{5/2}$-$3p_{3/2}$ &\hspace{5mm} 9.115 &\hspace{5mm}9.137  \\
\bottomrule\\
\end{tabular}
\end{table}

A measure of the quality of the calculations is also the comparison of the doublet member oscillator strength ratio, $\rho$=$f(3s$$\rightarrow$$np_{3/2})/f(3s$$\rightarrow$$np_{1/2})$, with available experiments.  This ratio can become significantly larger than the value of 2 -- expected in\begin{figure}[b]\centering\small\label{Na_3s(rho)}
\includegraphics[width=8.00cm]{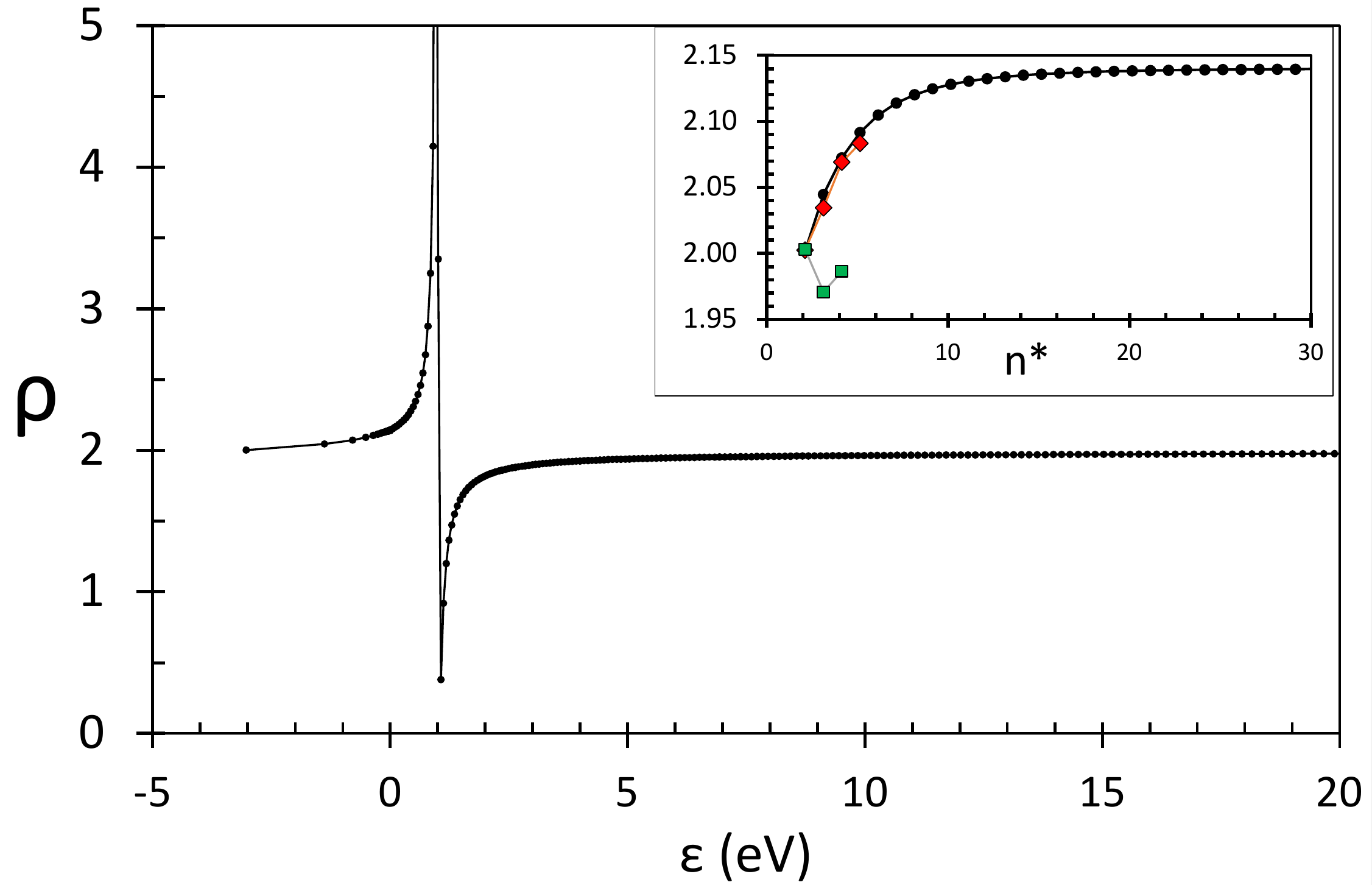}
\caption{\label{fig: Na_3s(rho)} (Color online) The ratio  $\rho=f(3s$$\rightarrow$$np_{3/2})/f(3s$$\rightarrow$$np_{1/2})$ for  Na $3s$$\rightarrow$$np$ transitions. 
Experimental: $\mdblksquare$ (green), data taken from Wiese et al.\ \cite{Wiese1969}. 
Theoretical: $\mdblkdiamond$ (red), using the reduced matrix elements of Safronova et al.\ \cite{Safronova1998};  $\mdblkcircle$ (black), present. }
\end{figure} the absence of spin-orbit interaction -- and is more visible in heavier alkali atoms \cite{Migdalek1998, CET2022b}.  Fig.\ \ref{fig: Na_3s(rho)} shows the value of $\rho$ for the first twenty n$p$ Rydberg states (insert) and its behavior as a function of electron energy as one traverses the ionization threshold and crosses the location of the minimum of the photoionization cross section.   We see that our results for the first four $p$ states agree well with the numbers extracted from the work of Safronova et al.\ \cite{Safronova1998}.  The experimental data extracted from Ref.\ \cite{Wiese1969} are quite off for 4$p$ and 5$p$. Even though the spin-orbit interaction effects are expected to be less significant for Na than for the heavier alkalis, this figure shows a significant departure from the value of 2 and a sharp discontinuity at the location of the Cooper minimum.  An easier, possibly, way to measure the location of this minimum could be to, instead, measure the ratio $\rho$, or the photoelectron angular distribution asymmetry parameter $\beta$ that becomes -1 at the minimum, from the typical value around 2.

\subsection{Na $4s$ state}
\begin{figure}[h]\centering\small\label{Na_4s(f)}
\includegraphics[width=8.0cm]{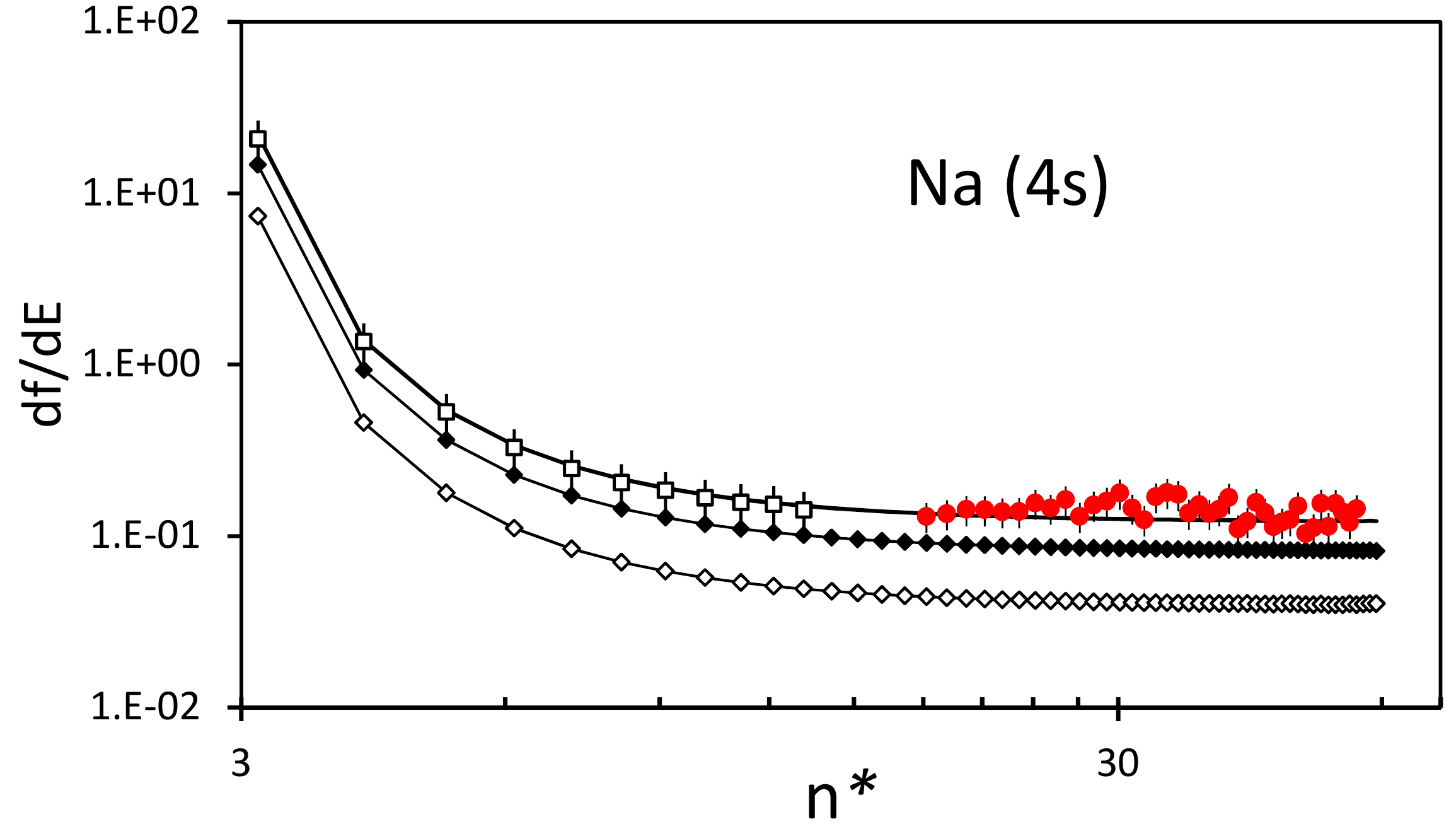}
\caption{\label{fig: Na_4s(f)}  (Color online) $df/dE$  for  Na $4s$$\rightarrow$$np$ transitions.
$\mdblksquare$, data taken from Wiese et al.\ \cite{Wiese1969}; $\mdwhtcircle$, Rafiq et al.\ \cite{Rafiq2008}; solid line, present $4s_{1/2}$$\rightarrow$$np_{3/2}+np_{1/2}$; $\mdblkdiamond$, present $4s_{1/2}$$\rightarrow$$np_{3/2}$; $\mdwhtdiamond$, present $4s_{1/2}$$\rightarrow$$np_{1/2}$}
\end{figure}

Rafiq et al.\ \cite{Rafiq2008} published data on the $4s$$\rightarrow$$np$ oscillator strengths that seem to be in fair agreement with theory, and serve as a smooth extension of the values from Wiese et al.\ \cite{Wiese1969} at lower $n$ (Fig.\ \ref{fig: Na_4s(f)}).  Fig.\ \ref{fig: Na_4s(f)} presents our results as compared with the data of Rafiq et al.\ \cite{Rafiq2008} and the older NBS data. The agreement between these data is satisfactory. There are no other published oscillator strength values available for comparison.  

\begin{figure}[h]\centering\small\label{Na_cs(4s)}
\includegraphics[width=8.59cm]{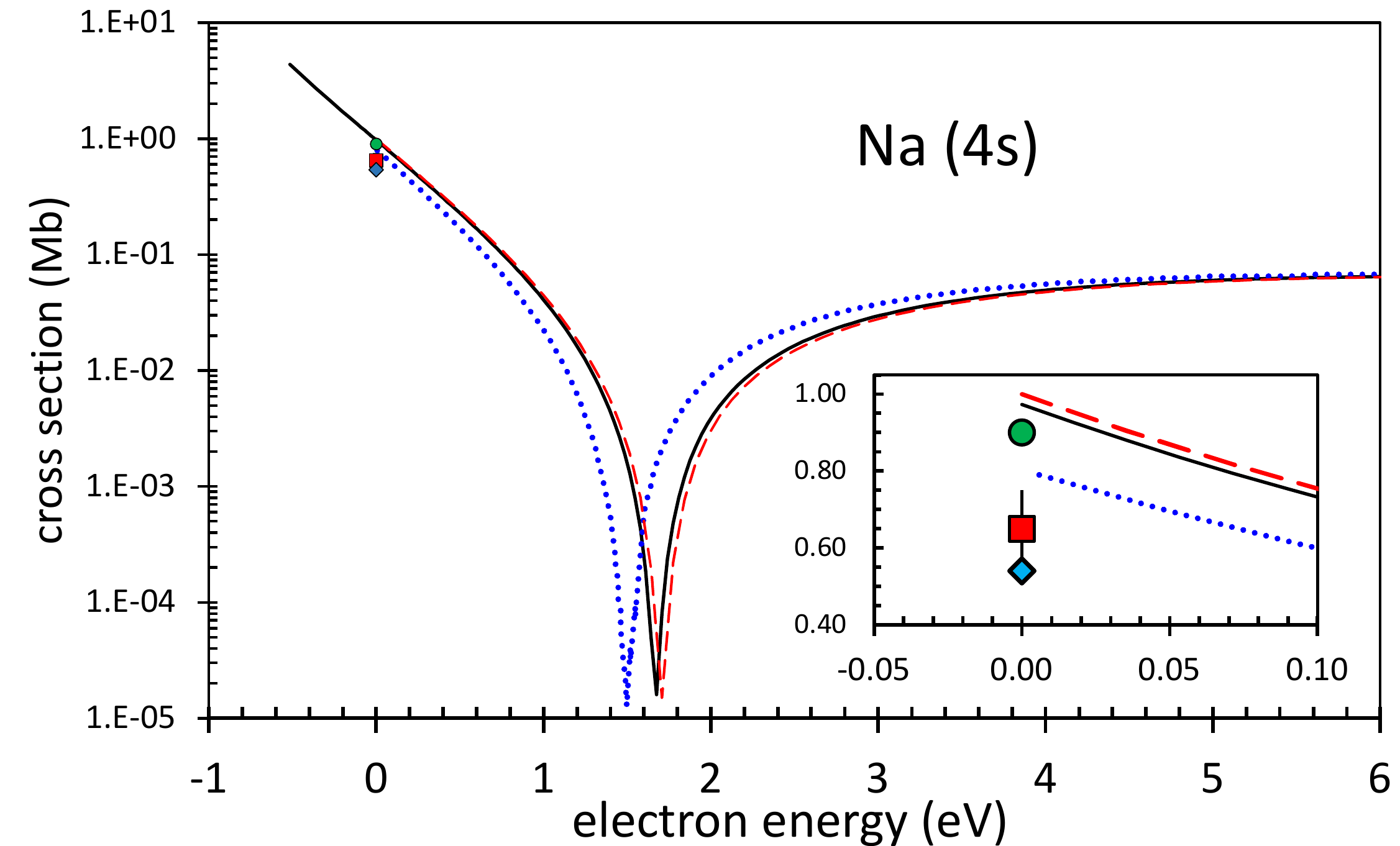}
\caption{\label{fig: Na_cs(4s)} (Color online)   Na $4s$ photoionization cross sections. Experimental: $\mdblksquare$ (red) Rafiq et al.\ \cite{Rafiq2008}. Theoretical: $\mdblkcircle$ (green), Moskvin \cite{Moskvin1963}; $\mdblkdiamond$ (blue), Aymar \cite{Aymar1978}; dotted line (blue), Singor et al.\ \cite{Singor2021}; solid line (black), present results; dashed line (red), present results using $a_d(\omega)$.   \textit{Insert}: enlargement of the threshold area.}
\end{figure}

For the photoionization cross sections there are the threshold estimates by Burgess and Seaton \cite{Burgess1960} and Moskvin \cite{Moskvin1963}, the older calculations by Aymar et al.\ \cite{Aymar1978}, predicting a Cooper minimum around the photoelectron energy of 2 eV, and more recently the fully relativistic treatment of Singor et al.\ \cite{Singor2021}.  Table III gives a comparison of all these results at threshold and Fig.\ \ref{fig: Na_cs(4s)} shows a comparison with the Singor et al.\ \cite{Singor2021} data.   We find an overall good agreement with Ref.\ \cite{Singor2021} although we predict slightly different locations for the Cooper minimum. The data of Aymar \cite{Aymar1978} at threshold are also in reasonable agreement.  Fig.\ \ref{fig: Na_cs(4s)} also shows the small effect of using $\alpha_d (\omega)$.
\begin{table}[h]\small\centering\label{Table3}
\caption{Comparison of the measured and calculated Na ($4s$) photoionization cross section value $\sigma_\text{th}$ (in Mb) at the ionization threshold.}
\begin{tabular}{cccccc}
\toprule\toprule
present& Ref.\cite{Rafiq2008}&Ref.\cite{Aymar1978}&Ref.\cite{Burgess1960}&Ref.\cite{Moskvin1963}&Ref.\cite{Singor2021}\\
\midrule
0.973&0.65(0.10)&$\sim$1.\footnote{extracted from the published graph}&0.90&0.90&0.79\\
\bottomrule
\end{tabular}
\end{table}

We are puzzled with the $\sigma(4s)=0.65(0.10)$ Mb value at threshold published by Rafiq et al.\ \cite{Rafiq2008}. Their optical oscillator strengths are in good agreement with our calculation, even as high as $n=57$; from those values we should project their measured cross section at threshold to have a value in the range of 0.9 Mb to 1.0 Mb, rather than 0.65 Mb, which coincides with our value for the \textit{partial} cross section $\sigma(4s$$\rightarrow$$\epsilon p_{3/2})$ at threshold.


\subsection{Na $3p$ state}
\begin{figure}[b]\centering\small\label{Na_3p(f)}
\includegraphics[width=8.0cm]{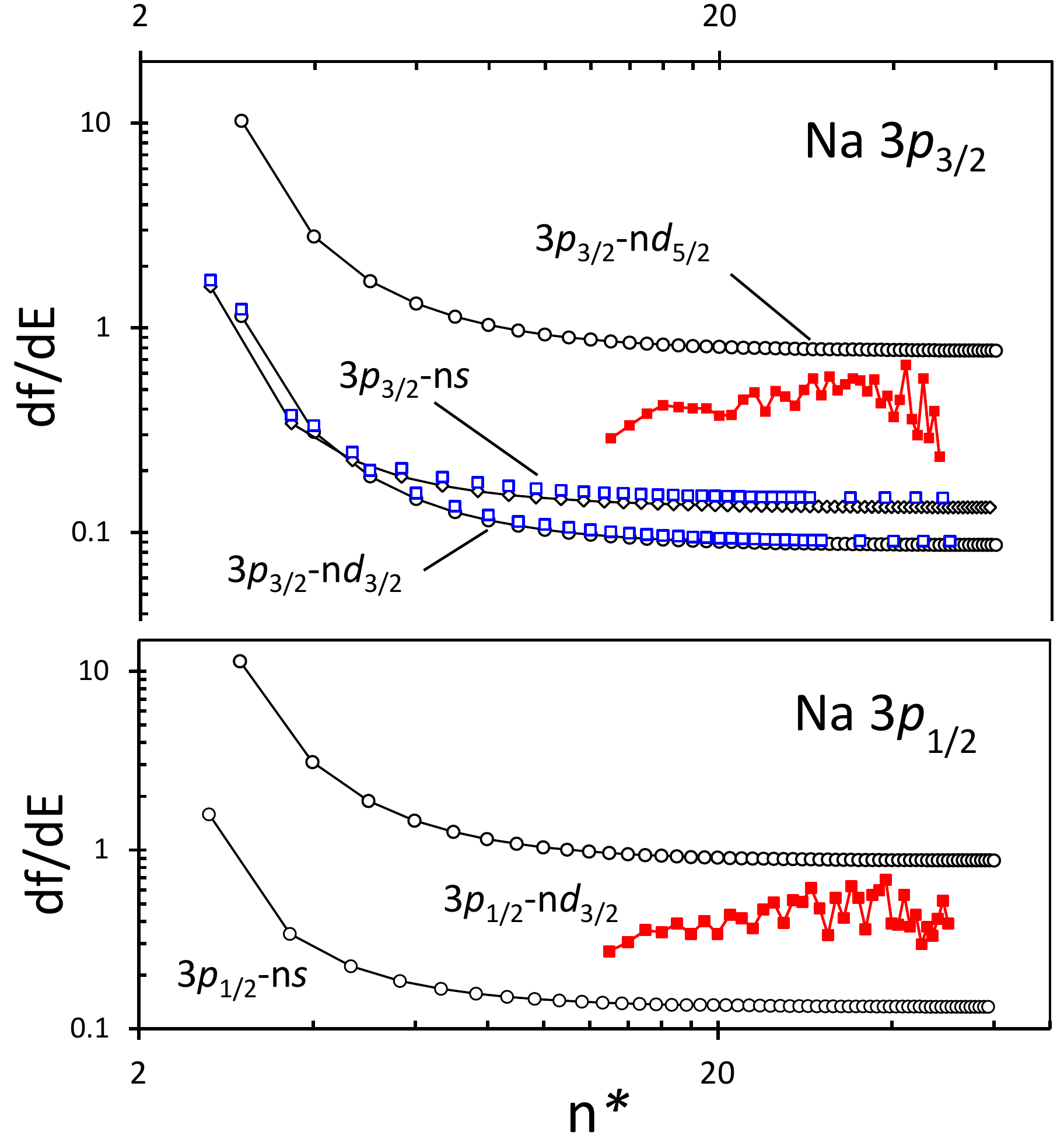}
\caption{\label{fig: Na_3p(f)} (Color online) $df/dE$  for Na $3p$$\rightarrow$$nd,ns$ transitions. 
\textit{Top:} Na $3p_{3/2}$. 
Experimental:  $\mdblksquare$, $3p_{3/2}$$\rightarrow$$nd_{5/2,3/2}$ (red), Baig et al.\ \cite{Baig2007};  
Theoretical: $\mdwhtsquare$, $3p_{3/2}$$\rightarrow$$nd_{3/2},ns$ (blue),  Miculis and Meyer \cite{Miculis2005}; $\mdwhtcircle$, present $3p_{3/2}$$\rightarrow$$nd_{5/2},nd_{3/2}$;  $\mdwhtdiamond$, present $3p_{3/2}$$\rightarrow$$ns$. 
\textit{Bottom:} Na $3p_{1/2}$. 
Experimental:  $\mdblksquare$, $3p_{1/2}$$\rightarrow$$nd_{3/2}$ (red), Baig et al.\ \cite{Baig2007}.  
Theoretical: $\mdwhtcircle$, present $3p_{1/2}$$\rightarrow$$nd_{3/2},ns$. }
\end{figure}

Fig.\ \ref{fig: Na_3p(f)} presents the absorption oscillator strengths for $3p$$\rightarrow$$nd$ and $ns$ transitions.  Our data have a smooth transition from $n=4$ to $n=80$, reaching the asymptotic behavior of $1/n^{*3}$. The data of Baig et al.\ \cite{Baig2007}  are not smooth and considerably lower than ours. 
The theoretical data of Miculis and Meyer \cite{Miculis2005} for  $3p_{3/2}$$\rightarrow$$nd_{3/2}$ transitions are essentially identical to ours on this Figure.
There is an ambiguity in the discussion of Ref.\ \cite{Baig2007}: since their measurements did not distinguish between the $d$-multiplet members $nd_{5/2}$ and $nd_{3/2}$ they compared their measurements with the ``average'' of the values published in Ref.\ cite{Miculis2005}.  Their data, in fact, need to be compared to the \textit{sum} of the doublet members.  
Our data for the $3p_{1/2}$$\rightarrow$$nd_{3/2}$oscillator strengths are not in good agreement with the measurements of Ref.\ \cite{Baig2007}, either in magnitude or in $n^*-$dependence. It should be noted that the measured $f-$values of  that work were calibrated to their measured photoinization cross sections at threshold, $\sigma_\text{thr}(3p_{3/2})=7.9(1.3)$ Mb and  $\sigma_\text{thr}(3p_{1/2})=6.7(1.1)$ Mb.  Our calculated values for these two quantities are 8.06 Mb and 8.11 Mb, respectively. 

For the photoionization cross sections we have for comparison several measurements \cite{Burkhardt1988, Baig2007, Wippel2001, Preses1985, Rothe1969} made over the years. The energy dependence of the cross section is smooth.  Our calculations (Fig.\ \ref{fig: Na_cs(3p)}) are in excellent agreement with the experimental results of Burkhardt et al.\ \cite{Burkhardt1988}, Preses et al.\ \cite{Preses1985}, Baig et al.\ \cite{Baig2007}, Rothe \cite{Rothe1969}, and Wippel \cite{Wippel2001}. 
Our results are also in very good agreement with the theoretical ones of Petrov et al.\ \cite{Petrov2000} and of Singor et al.\ \cite{Singor2021} over the entire energy range they mutually cover; they are indistinguishable on the scale of Fig.\ \ref{fig: Na_cs(3p)}.  The theoretical values of Miculis and Meyer \cite{Miculis2005} are slightly lower but vey close to ours; they are not shown in this {(busy) graph.
\begin{figure}[h]\centering\small\label{Na_cs(3p)}
\includegraphics[width=8.7cm]{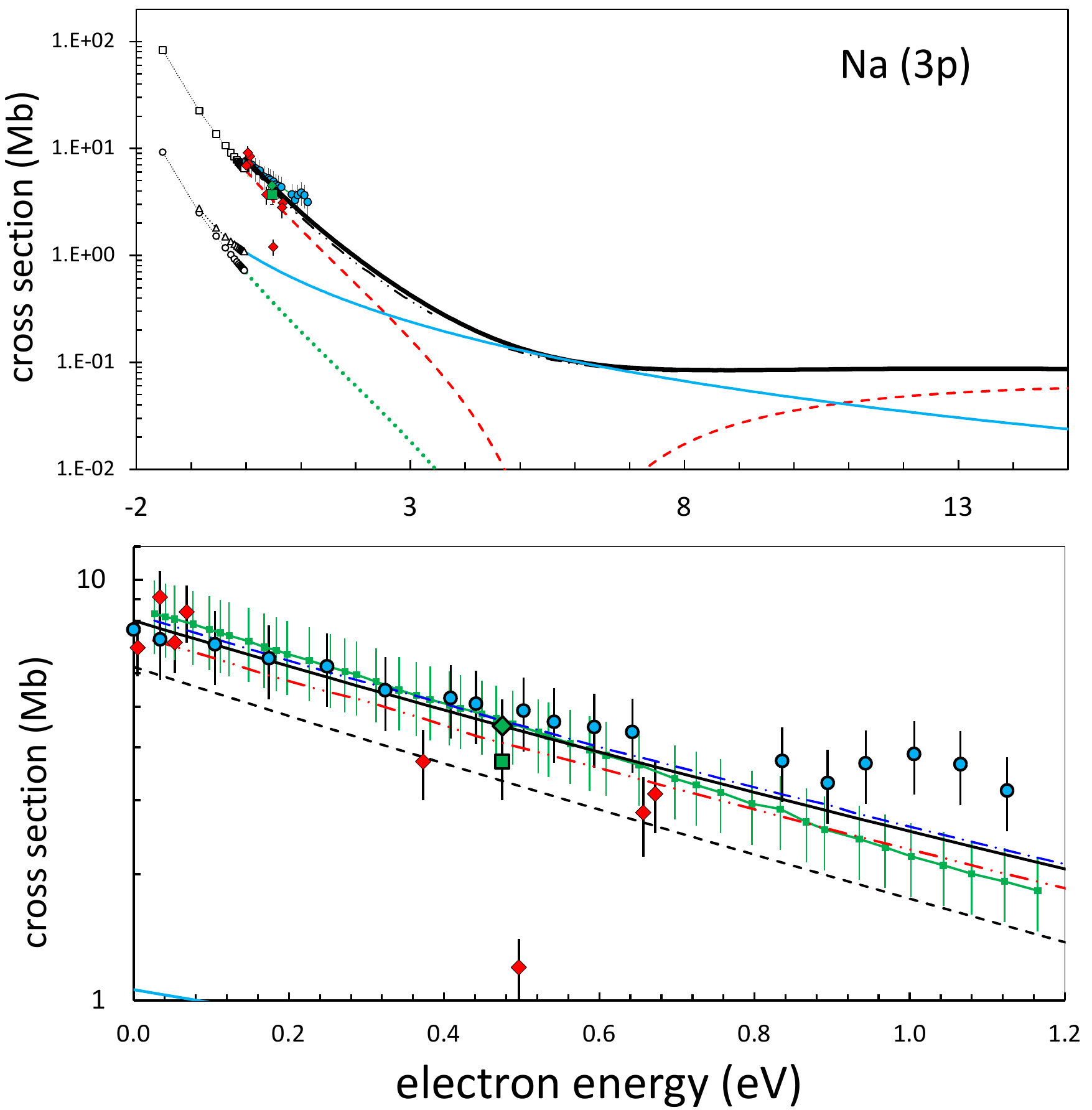}
\caption{\label{fig: Na_cs(3p)} (Color online) Na $3p$  photoionization cross sections.
Experimental: $\mdblkdiamond$, Wippel et al.\ \cite{Wippel2001};  $\mdwhtsquare$,$\mdwhtdiamond$, Burkhardt et al.\ \cite{Burkhardt1988}; 
$\mdwhtcircle$, Rothe \cite{Rothe1969}; small $\mdblksquare$ connected with a line, Preses et al.\ \cite{Preses1985}.
Theoretical: dash-dot, Petrov et al.\ \cite{Petrov2000};  dash-double dot, Singor et al.\ \cite{Singor2021}; dotted line, present $3p_{3/2}$$\rightarrow$$\epsilon d_{5/2}$; dashed line, present $3p_{3/2}$$\rightarrow$$\epsilon d_{3/2}$; thin black line, present $3p_{3/2}$$\rightarrow$$\epsilon s_{1/2}$; thick black line, present, total photoionization cross section of $3p_{3/2}$. \textit{Lower panel}: enlargement of the threshold area.}
\end{figure}

We note that our \textit{partial} cross section $\sigma(3p$$\rightarrow$$\epsilon{}d)$ has a deep Cooper minimum that is masked by the $\sigma(3p$$\rightarrow$$\epsilon s)$ partial cross section.  The location of the minimum can be experimentally confirmed by use of a circularly polarized laser source to separate the $\epsilon{}f$ partial wave.  Such measurements were made early on by  Duong et al.\ \cite{Duong1978}, but were limited to photoelectron energies less than 0.5 eV.  Yet, they provide additional salient features of the 3$p$ photoionization process and are important to compare with theory.  The experiments involved two sources of polarized light and measured the spin-polarized electron intensities.  The measured quantities are related to the partial photoionization cross sections \cite{Duong1978}
\begin{align}
S_{+}(\epsilon)=& \frac{k}{5} \sigma_{3p,\epsilon d}(\epsilon) \\
S_{-}(\epsilon)=&\frac{k}{3}\left [ 0.1{\ }\sigma_{3p,\epsilon d}(\epsilon)+ \sigma_{3p,\epsilon s}(\epsilon) \right] \\
S(\epsilon)=S_{+}(\epsilon)+S_{-}(\epsilon)=&\frac{k}{3} \left [ 0.7{\ }\sigma_{3p,\epsilon d}(\epsilon) + \sigma_{3p,\epsilon s}(\epsilon) \right]\\
\rho(\epsilon)=\frac{S_{+}}{S_{-}}=&\frac{0.6{\ }\sigma_{3p,\epsilon d}(\epsilon)}{ 0.1{\ }\sigma_{3p,\epsilon d}(\epsilon) + \sigma_{3p,\epsilon s}(\epsilon)}
\end{align}
where $k$ is a normalization constant.
\begin{figure}[b]\centering\small\label{Na_(3p)_S_rho_c}
\includegraphics[width=6.7cm]{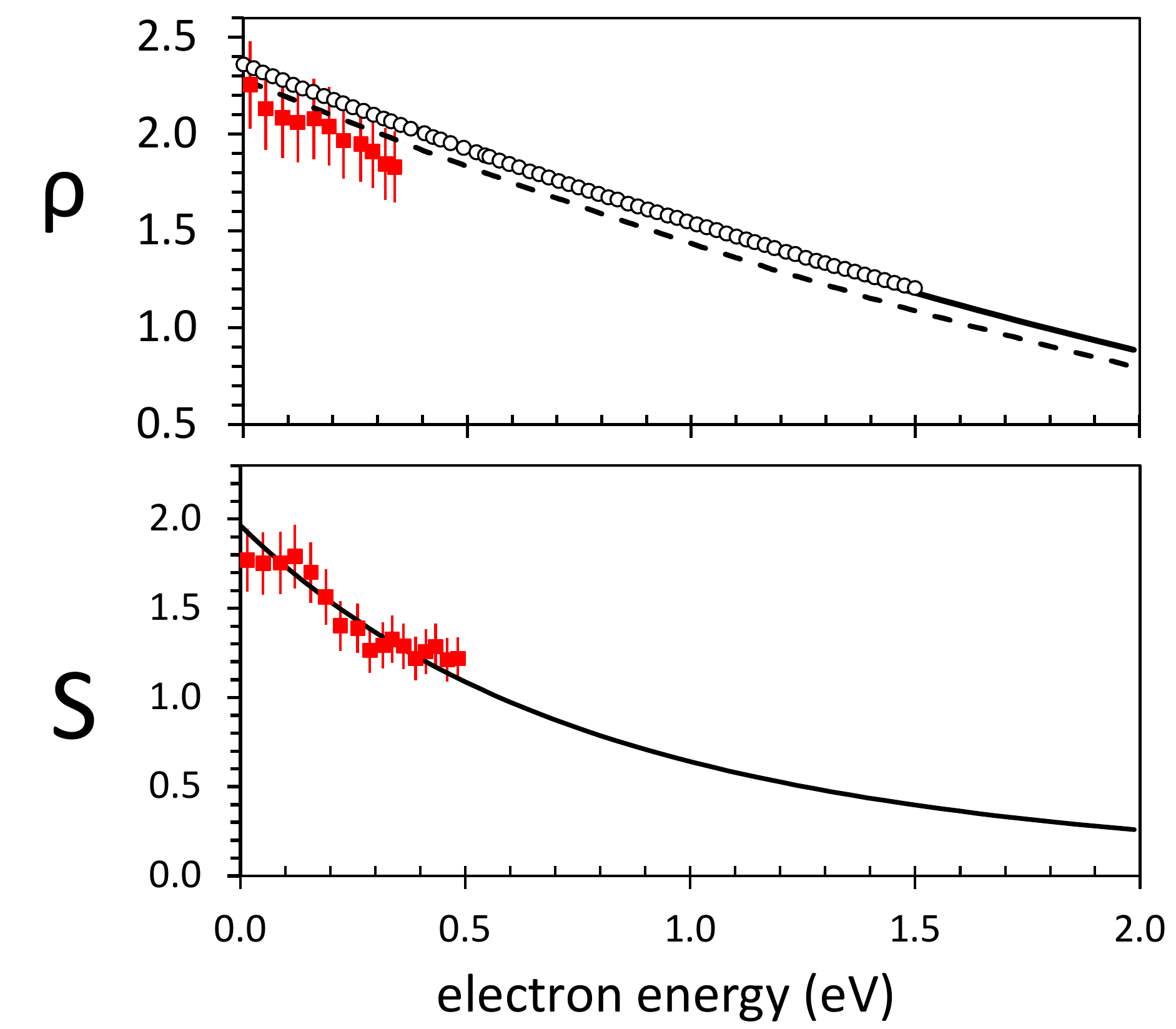}
\caption{\label{fig: Na_(3p)_S_rho_c} (Color online) Na $3p$  values for $\rho(\epsilon)$ and $S(\epsilon)$ for experiments with circularly polarized light.
Experimental: $\mdblksquare$ (red), Duong et al.\ \cite{Duong1978};  
Theoretical: $\mdwhtcircle$, Petrov et al.\ \cite{Petrov2000};  dashed line, Miculis and Meyer \cite{Miculis2005}; solid line (black), present.}
\end{figure}

Fig.\ \ref{fig: Na_(3p)_S_rho_c} presents the results of the measurements of $S(\epsilon)$ and $\rho(\epsilon)$.  We see a nice agreement of the measured values with our calculations for both these quantities.  The \textit{relative} experimental values of $S$ were brought to absolute scale by normalizing them to our results.  There are no other available calculations for $S$ to compare them with.  
The measured values of $\rho$ follow a very similar trend with our calculations as well as those of Refs.\ \cite{Petrov2000, Miculis2005}.  The data of Miculis and Meyer \cite{Miculis2005} seem to agree closer in absolute value, but the stated experimental error of 5\%-10\% overlaps with all three calculations.  Our data are ``embarrassingly'' close to those of Petrov et al.\ \cite{Petrov2000}; they are indistinguishable from them on the graph.
It would be very interesting in the future to have similar measurements, i.e. that separate the partial photoionization cross sections, extended to higher photon energies, covering the $\sigma (3p$$\rightarrow$$\epsilon{}d)$ partial cross section minimum around $\epsilon=6$ eV.

\subsection{Na  $3d$ state}
Fig.\ \ref{fig: Na_cs(3d)} presents our calculated $3d$$\rightarrow$$\epsilon f$ cross sections along with with a single experimental value available from Nadeem et al.\ \cite{Nadeem2013} at threshold.  The agreement is satisfactory.
\begin{figure}[h]\centering\small\label{Na_cs(3d)}
\includegraphics[width=8.00cm]{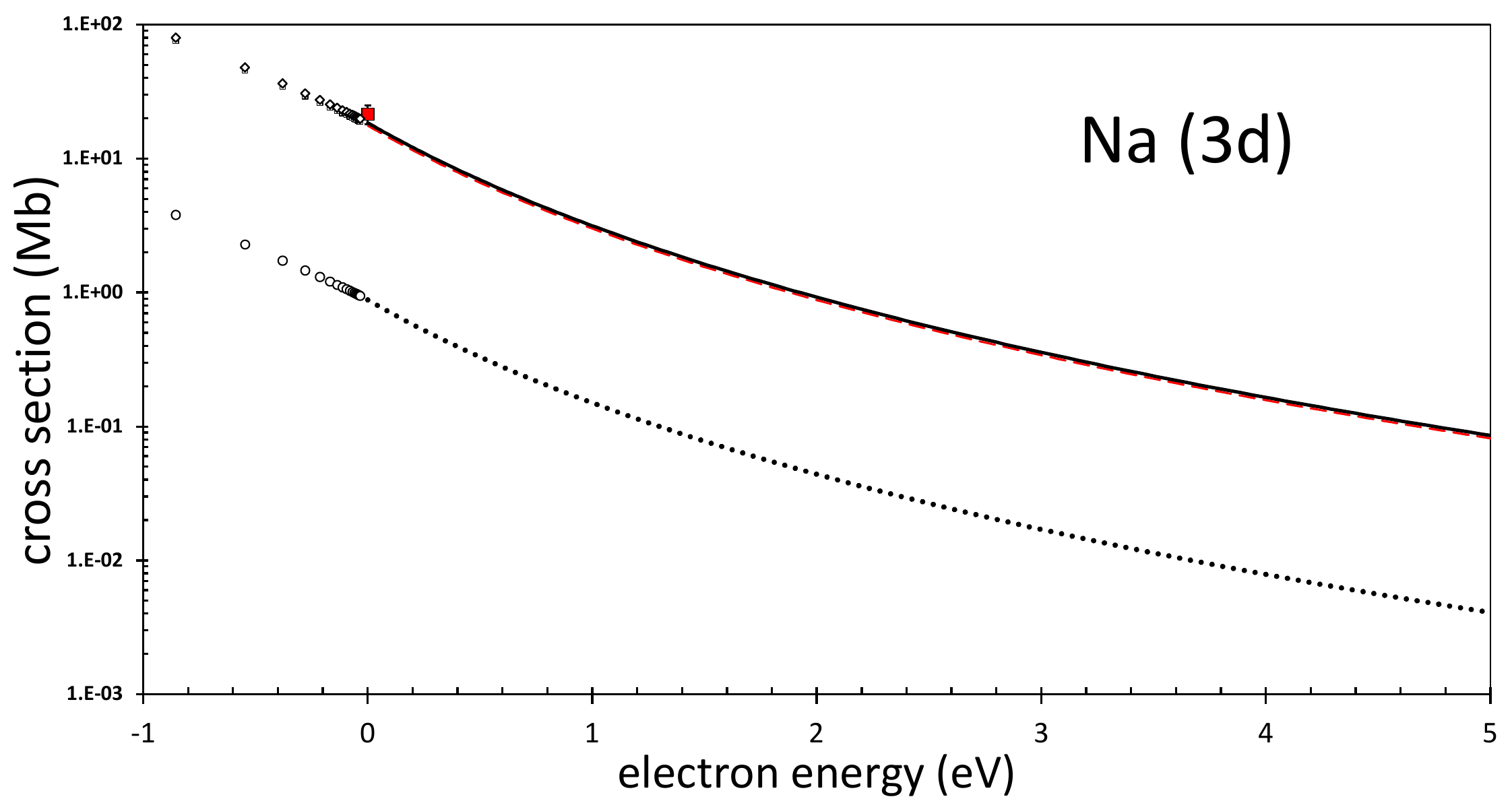}
\caption{\label{fig: Na_cs(3d)} (Color online) Na($3d$) photoionization cross sections.
Experimental: $\mdblksquare$, $3d_{5/2}$, Nadeem et al.\ \cite{Nadeem2013}.
Theoretical: solid line, present $3d_{5/2}$$\rightarrow$$\epsilon f_{5/2}+\epsilon f_{7/2}$; dotted line, present $3d_{5/2}$$\rightarrow$$\epsilon f_{5/2}$; dashed line (red), present $3d_{5/2}$$\rightarrow$$\epsilon f_{7/2}$.  The data of Singor et al.\ \cite{Singor2021} are indistinguishable from ours on this graph.}
\end{figure}

Nadeem et al.\ \cite{Nadeem2015} also presented data and discussion of the $3d$$\rightarrow$$nf$ absorption oscillator strengths. For comparison, we examined the transitions to $nf$ and $np$ states with $n=4-80$. Our results are indicated in Fig.\ \ref{fig: Na_cs(3d)} for a few discreet transitions as values for ``negative energy,'' to show the smooth transition between discrete and continuum photoabsorption. Fig.\ \ref{fig:  Na_f(3d-nf)_c} presents the calculated optical oscillator strengths for all transitions, in the form of $df/dE$, and compares them with the data of Nadeem et al.\ \cite{Nadeem2015}, along with the very early tabulation of Wiese et al.\ \cite{Wiese1969} as a standard reference.  Our data have a smooth transition from $n=4$ to $n=80$, reaching the asymptotic bahavior of $1/n^{*3}$.  The agreement with the Wiese  data \cite{Wiese1969} is excellent.  There is overall a fair agreement with the measurements of Nadeem et al.\ \cite{Nadeem2015}, though their data are not smooth in their $n^*-$dependence.
\begin{figure}[h]\centering\small\label{Na_f(3d-nf)_c}
\includegraphics[width=6.5cm]{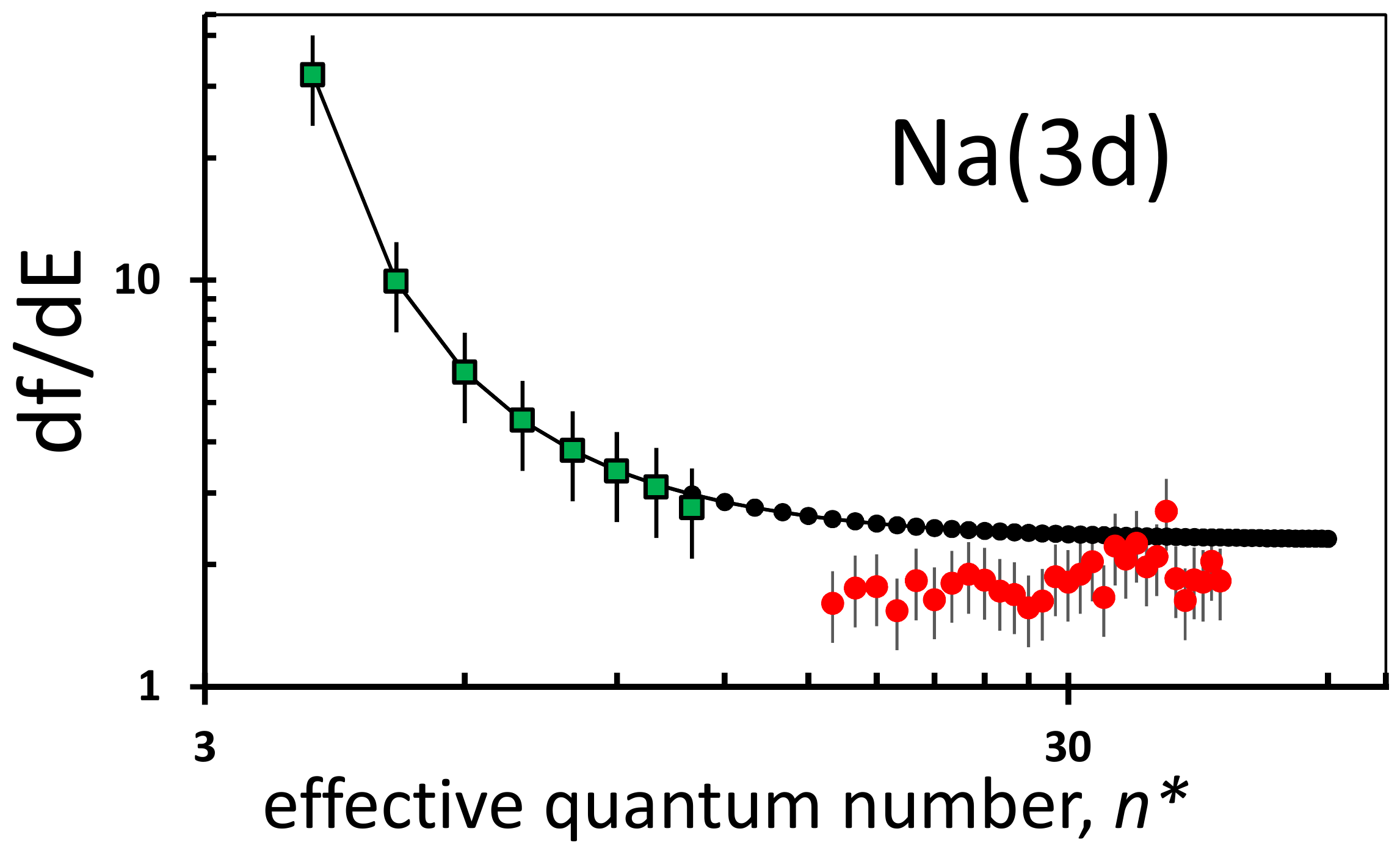}
\caption{\label{fig: Na_f(3d-nf)_c} 
(Color online) Na $3d$$\rightarrow$$nf_{5/2,7/2}$ absorption oscillator strengths.  
Experimental: $\mdwhtsquare$ (green), data taken from Wiese et al.\ \cite{Wiese1969}; $\mdwhtcircle$ (red), $3d$$\rightarrow$$nf$, Nadeem et al.\ \cite{Nadeem2015}.  
Theoretical: $\mdblkcircle$, present; the $nd_{5/2}$ and $nd_{3/2}$ values are indistinguishable on the graph.}
\end{figure}

In view of the above comparisons of various works with the results of Baig et al.\ \cite{Baig2007} and Nadeem et al.\ \cite{Nadeem2015}, and the latter's modest agreement with theory, concerns can be raised about the overall accuracy of these data for optical oscillator strengths and they need to be reevaluated.

\vfill%
\pagebreak

%




\begin{thebibliography}{}
\bibitem{Cooper1962} J. W. Cooper, Phys. Rev. {\bf 128}, 681 (1962) 
\bibitem{Fano1968} U. Fano and J. W. Cooper, Rev. Mod. Phys. {\bf 40}, 441 (1968)
\bibitem{Seaton1951}M. J. Seaton, Proc. R. Soc. Loandon Ser. A {\bf 208}, 218 (1951)
\bibitem{Ditchburn1953} R. W. Ditchburn, P. J. Jutsum, and G. V. Marr, Proc. R. Soc. London Ser. A {\bf 219}, 89 (1953).
\bibitem{Hudson1965} R. D. Hudson and V. L. Carter, Phys. Rev. {\bf 139} A1426 (1965)
\bibitem{Sandner1981} W. Sandner, T. F. Gallagher, K. A. Safinya, and F. Gounand, Phys. Rev., A {\bf 23}, 2732 (1981)
\bibitem{Chang1975}T. N. Chang, J. Phys. B {\bf 8}, 743 (1975)
\bibitem{Saha1989}H. P. Saha, Phys. Rev., A {\bf 39}, 628 (1989)
\bibitem{Fink1986}M. G. J. Fink and W. R. Johnson, Phys. Rev., A {\bf 34}, 3754 (1986)
\bibitem{Weisheit1972}J. C. Weisheit, Phys. Rev., A {\bf 5}, 1621 (1972), and references  therein
\bibitem{Marinescu1994} M. Marinescu, H. R. Sadeghpour, and A. Dalgarno, Phys. Rev., A {\bf 49}, 982 (1994)
\bibitem{Petrov1999} I. D. Petrov, V. L. Sukhorukov, and H. Hotop, J. Phys. B: At. Mol. Opt. Phys. {\bf 32} 973 (1999)
\bibitem{Petrov2000}I. D. Petrov, V. L. Sukhorukov, E. Leber, and H. Hotop, Eur. Phys. J. D {\bf 10}, 53 (2000).
\bibitem{Zatsarinny2010} O. Zatsarinny and S. S. Tayal, Phys. Rev. A {\bf 81}, 043423 (2010)
\bibitem{Singor2021}A. Singor, D. Fursa, I. Bray, and R. McEachran, Atoms, {\bf 9}, 42 (2021)
\bibitem{Amin2008}N. Amin, S. Mahmood, S. U. Haq, M. A. Kalyar, M. Rafiq, and M. A. Baig, J. Quant. Spectrosc. Radiat. Transfer {\bf 109}, 863 (2008) 
\bibitem{Yar2013} A. Yar, R. Ali, and M. Aslam Baig ,  Phys. Rev. A {\bf88}, 033405 (2013)
\bibitem{Kalyar2016} M. A. Kalyar, A. Yar, J. Iqbal,R. Ali, and M. A. Baig, J. Opt. Las. Tec. {\bf 77} 72 (2016)
\bibitem{Rafiq2008}M. Rafiq, M.A. Kalyar, and M.A. Baig, J. Phys. B: Atom. Mol. Opt. Phys. {\bf 41} 115701 (2008) 
\bibitem{Baig2007}M. A. Baig, S. Mahmood,  M. A. Kalyar, M. Rafiq,N. Amin, and S. U. Haq,  Eur. J. Phys. D {\bf 44}, 9-16 (2007) 
\bibitem{Safronova1998}M. S. Safronova, A. Derevianko, and W.R. Johnson, Phys. Rev. A {\bf 58}, 1016 (1998)
\bibitem{CET1984a} C. E. Theodosiou, Phys. Rev. A {\bf 30}, 2881 (1984)
\bibitem{CET1984b} C. E. Theodosiou, Phys. Rev. A {\bf 30}, 2910 (1984)
\bibitem{Curtis1993} L. J. Curtis, Phys. Scripta, {\bf 48}, 559 (1993)
\bibitem{CET1996} C. E. Theodosiou, L. J. Curtis, and C. A. Nicolaides, Phys. Rev., A{\bf 52}, 3677 (1995)
\bibitem{CET2022a} C. E. Theodosiou, arXiv:2211.07831
; Phys. Rev. A (submitted) (2022)
\bibitem{CET2022b} C. E. Theodosiou, arXiv:http://arxiv.org/abs/2211.12664
; J. Phys. B: Atom. Mol. Opt. Phys. (submitted) (2022)
\bibitem{Desclaux1969} J. P. Desclaux, Comput. Phys. Commun., {\bf 1}, 216 (1969)
\bibitem{Johnson1981} W. R. Johnson, D. Kolb, and K.-N. Huang, Atom. Data Nucl. Data Tables, {\bf 28}, 333 (1983)
\bibitem{Norcross1973} D. W. Norcross, Phys. Rev. A {\bf 7}, 606 (1973)
\bibitem{Sobelman1992}I. I. Sobelman, \textit{Atomic Spectra and Radiative Transitions}, Second Edition, Springer (1992)
\bibitem{Konovalov2016}A. V. Konovalov and A. N. Ipatov, http://dx.doi.org/10.1016/ j.spjpm.2016.02.08  (2016)
\bibitem{Wiese1969}W.L. Wiese, M.W. Smith, B.M. Miles, \textit{Atomic Transition Probabilities}, {\bf 22}, NSRDS-NBS, New York, USA, 1969
\bibitem{Nawaz1992}M. Nawaz, W.A. Farooq, and J.-P. Connerade, J. Phys. B: At. Mol. Opt. Phys. {\bf 25} 5327 (1992).
\bibitem{McEachran1969}R. P. McEachran, C.E. Tull, and M. Cohen, Can. J. Phys {\bf 47}, 835 (1969)
\bibitem{Migdalek1998}J. Migdalek and Y.-K. Kim, J. Phys. B: Atom. Mol. Opt. Phys. {\bf 31} 1947 (1998) 
\bibitem{Burgess1960}A. Burgess and M. J. Seaton, Mon. Not. R. Astron. Soc. {\bf 120} 1213 (1960) 
\bibitem{Moskvin1963}Y V Moskvin, Opt. Spectrosc. {\bf 15}, 316 (1963) 
\bibitem{Aymar1978}M. Aymar,  J. Phys. B: Atom. Mol. Opt. Phys. {\bf 11} 1413 (1978) 
\bibitem{Miculis2005}K. Miculis and W. Meyer, J. Phys. B: Atom. Mol. Opt. Phys. {\bf 38}2097-2108 (2005)
\bibitem{Burkhardt1988} C.E. Burkhardt, J. L. Libbert, Jian Xu, J.J. Leventhal, and J.D. Kelley, Phys. Rev. A {\bf 38}, 5949 (1988)
\bibitem{Rothe1969}D E J Rothe, J. Quant. Spectrosc. Radiat. Transfer {\bf 9}, 49 (1969)
\bibitem{Preses1985} J.M. Preses, C.E. Burkhardt, R.L. Corey, D.L. Earson, T.L. Daulton, W.P. Garver, J.J. Leventhal, A.Z. Msezane, and S.T. Manson, Phys. Rev. A {\bf 32}, 1264 (1985)
\bibitem{Wippel2001} V. Wippel, C. Binder, W. Huber, L. Windholz, M. Allegrini, F. Fuso, and E. Arimondo, Eur. J. Phys. D {\bf 17}, 285-291 (2001)
\bibitem{Duong1978}H. T. Duong, J. Pinard, and J.-L. Vialle, J. Phys. B: Atom. Mol. Opt. Phys. {\bf 11} 797 (1978) [3p]
\bibitem{Nadeem2013}A. Nadeem, M. Shah, S. Shahzada, M. Ahmed, and S. U. Haq, Eur. J. Phys. D, {\bf 67} 196 (2013) 
\bibitem{Nadeem2015}A. Nadeem, M. Shah, S. Shahzada, M. Ahmed, and S. U. Haq, J. Appl. Spectr., {\bf 82} 659 (2015) 
\end{thebibliography}
\end{document}